\documentclass{ws-ijmpe}

\usepackage{bm}

\newcommand{\be}{\begin{equation}}
\newcommand{\ee}{\end{equation}}
\newcommand{\bn}{\begin{eqnarray}}
\newcommand{\en}{\end{eqnarray}}
\newcommand{\ba}{\begin{array}}
\newcommand{\ea}{\end{array}}
\newcommand{\bc}{\begin{center}}
\newcommand{\ec}{\end{center}}
\newcommand{\bml}{\begin{mathletters}}
\newcommand{\eml}{\end{mathletters}}

\newcommand{\shf}{{{\sc shf}}}

\newcommand{\ld}{{{\sc ld}}}
\newcommand{\nse}{{{\sc nse}}}

\begin{document}

\markboth{M. Rafalski, W. Satu{\l}a, R.A. Wyss }{Mass number
dependence of the Skyrme-force-induced nuclear symmetry energy}

%
\catchline{}{}{}{}{}
%

\author{\footnotesize M. RAFALSKI\footnote{rafalski@fuw.edu.pl}}
\address{Institute of Theoretical Physics, University of Warsaw, \\
ul. Ho\.za 69, 00-681 Warsaw, Poland\\
}

\author{W. SATU{\L}A\footnote{satula@fuw.edu.pl}}
\address{Institute of Theoretical Physics, University of Warsaw, \\
ul. Ho\.za 69, 00-681 Warsaw, Poland,\\
KTH (Royal Institute of Technology)\\
AlbaNova University Center, 106 91 Stockholm, Sweden\\
}

\author{R.A. WYSS\footnote{wyss@kth.se}}
\address{KTH (Royal Institute of Technology)\\
AlbaNova University Center, 106 91 Stockholm, Sweden\\
}

\title{MASS NUMBER DEPENDENCE OF THE SKYRME-FORCE-INDUCED NUCLEAR
SYMMETRY ENERGY}

\maketitle
\begin{history}
\received{(received date )}
\revised{(revised date )}
\end{history}

\begin{abstract}
The  global mass dependence of the nuclear symmetry energy
$a_{sym}(A)$ and its two basic ingredients due to the mean-level spacing
$\varepsilon (A)$ and effective strength of the isovector mean-potential
$\kappa (A)$ is studied within the Skyrme-Hartree-Fock model.
In particular, our study
determines the ratio of the surface-to-volume contributions to
$a_{sym}(A)$ to be $r_{S/V}\sim 1.6$ and reveals that
after removing momentum-dependent effects by rescaling
$\varepsilon$ and  $\kappa$ with isoscalar and isovector effective
masses, respectively, one obtains
$\varepsilon^\star \approx \kappa^\star$.
\end{abstract}

\section{Introduction}

The most common route in constructing effective microscopic nuclear models
starts from the nuclear equation of state in infinite nuclear
matter and use such parameters like isoscalar saturation density
$\rho_0$, the volume binding energy $a_V$, the incompressibility
parameter $K_\infty$, and the asymmetry energy $a_{sym}^{(V)}$
as primary constraints for these models. All these values are
extrapolated, however, from the studies of finite nuclei, in particular from
the semi-empirical mass formulas:
\be\label{ldrop} \frac{E}{A} = -a_V +
\frac{a_S}{A^{1/3}}+  \ldots + \left[
a_{sym}^{(V)} - \frac{a_{sym}^{(S)}}{A^{1/3}} +\ldots \right]
\left( I^2+ \lambda\frac{I}{A} \right) + \ldots ,
\ee
where $I\equiv |N-Z|/A$ while $a_S$ and  $a_{sym}^{(S)}$ are coefficients
defining contributions from the surface energy and the surface part of the
nuclear symmetry energy ({\sc nse}), respectively.

Concerning the isovector terms in finite nuclei which are subject of this
work there is at present no consensus concerning the magnitude, $\lambda$,
as well as origin of the Wigner, $\sim$$I$, term.
The Weizs\"acker mass formula ({\sc ld})
sets $\lambda \equiv 0$. The {\sc ld} of
M\"oller~{\it et al.}~\cite{[Mol95]} admits only a volume-type Wigner term,
which is not consistent with Eq.~(\ref{ldrop}). For this  term
it gives $\lambda \approx 0.975$ i.e. a value
close to the one used in
shell-model inspired mass formulas which
are using $\sim T(T+1)$ (i.e. $\lambda \equiv 1$) where $T=|T_z|=|N-Z|/2$.
 Another controversy exists concerning the surface contribution
 to the~\nse. The values of the surface-to-volume ratio $r_{S/V} =
 a_{sym}^{(S)} / a_{sym}^{(V)}$ quoted in the literature vary
 strongly. For example,
 Danielewicz~\cite{[Dan03]} estimates it to be
 $ 2.0 \leq r_{S/V} \leq 2.8$, the mass formula
 of Ref.~\cite{[Mol95]} yields $r_{S/V} \approx 1.6 $ while
 the hydrodynamical-type models that include properly
 polarization of the isovector density predict
 $r_{S/V} \approx 2 $~\cite{[Lip82]}.

The main objective  of this work is to study various aspects
of the~\nse~emerging within the Skyrme-Hartree-Fock (\shf)~model.
This work supplements our earlier studies on this
subject~\cite{[Sat03],[Sat05a],[Ban05],[Sat05c]} to which
we refer our readers for details.

\section{The Skyrme-force induced nuclear symmetry energy and its mass
number dependence}

 In Ref.~\cite{[Sat03]} we have demonstrated that the~\shf~symmetry
energy behaves rather unexpectedly according to the formula:
\be\label{esym}
   E_{sym}^{(SHF)} = \frac{1}{2} \varepsilon (A,T_z) T^2 +
                     \frac{1}{2} \kappa (A,T_z)  T(T+1) \, ,
\ee
where $\varepsilon (A,T_z)\approx \varepsilon(A) $ and
$\kappa (A,T_z) \approx \kappa (A)$ are fairly independent on
$T_z$, at least for $T_z \geq 8$.
Hence, for further quantitative analysis of
the mass dependence of the~\nse~we use the mean values of
${\bar{\varepsilon}}(A)$ and ${\bar{\kappa}} (A)$.
These averages over $T_z$ at fixed
$A$ are calculated using the following restricted set of
nuclei: $T_z\geq 4$ for $A =20$;  $T_z\geq 6$ for $A=24$; and $T_z\geq 8$ for
$28 \leq A\leq 128$. By using a restricted set of nuclei we smooth out
both  $\bar{\varepsilon}(A)$ and $\bar{\kappa} (A)$ curves
in order to diminish the influence of shell structure.
The ${\bar{\varepsilon}}(A)$ and ${\bar{\kappa}} (A)$ are interpreted as
the mean-level spacing at the Fermi energy in {\it iso-symmetric nucleus\/} and
the effective strength of the isovector
mean-potential, respectively.

\begin{figure}[t]
\par
\centerline{\epsfig{file=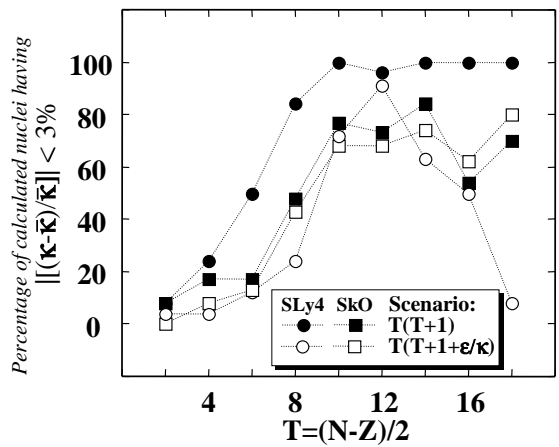,height=6cm,clip=}}
\vspace*{8pt}
\caption{Values of $\kappa (A,T_z)$ calculated using Sly4 (circles) and
SkO (squares). The calculated points mark the
percentage of calculated nuclei for
which $\kappa (A,T_z)$  deviates from the mean-value $\bar{\kappa} (A)$
by less than three percent versus $T_z$. Two different scenarios
are assumed for the linear contribution to $\sim \kappa T(T+\lambda)$
term with $\lambda =1$ (filled symbols) and $\lambda (\equiv \lambda_{RMF}) =
1+ {\bar{\varepsilon}}/{\bar{\kappa}}$ (open symbols).  Note that
while for the SLy4 $\lambda =1$, for the SkO it is half way between
$1< \lambda < \lambda_{RMF}$.
}\label{fig1}
\end{figure}

The relation~(\ref{esym}) holds extremely well within the~\shf~model
except for the SkO parameterization~\cite{[Rei99]}
for which we observe an enhancement
in the linear part of the second term in Eq.~(\ref{esym}). For the SkO
the second term  becomes $\sim \kappa T(T+\lambda)$ with $\lambda > 1$, as
illustrated in Fig.~\ref{fig1}. The value of $\lambda$ is however
much weaker than
$\lambda_{RMF} \approx 1+ {\bar{\varepsilon}}/{\bar{\kappa}}$
found  recently in relativistic mean-field ({\sc rmf})~\cite{[Ban05]}.
It should be mentioned that the SkO is characterized by an exceptionally
strong isovector strength of the spin-orbit term, inspired by
{\sc rmf}, having opposite sign
compared to standard {\sc sf} parametrization.

\begin{figure}[h]
\par
\centerline{\epsfig{file=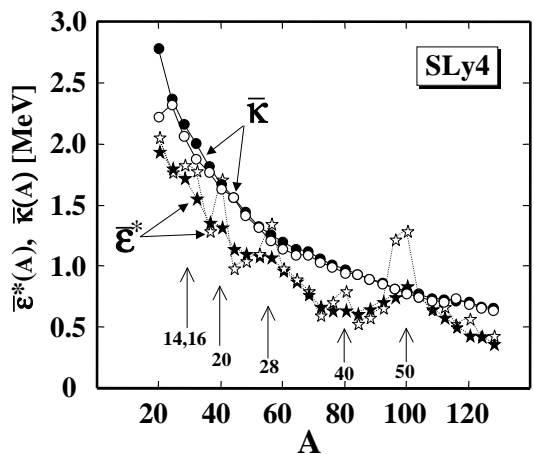,height=6cm,clip=}}

\caption[]{The isoscalar effective mass scaled values of
${\bar\varepsilon}^\star (A) \equiv  \frac{m_0^\star}{m} {\bar\varepsilon}(A)$
(stars) and ${\bar\kappa} (A)$ (circles) calculated using the~\shf~method with
SLy4 parametrization. Open symbols denote ${\bar\varepsilon}^\star (A)$ and
${\bar\kappa}(A)$ averaged over all the calculated nuclei. Filled symbols mark
smoothed values of ${\bar\varepsilon}^\star (A)$ and ${\bar\kappa} (A)$
calculated using a restricted set of data. Vertical arrows indicate major shell
gaps. Note the strong influence of shell structure on ${\bar\varepsilon}^\star
(A)$ and the smooth behavior of ${\bar\kappa}(A)$. } \label{sl4}
\end{figure}

The global mass dependence
of the two components of the symmetry energy, $\bar\varepsilon$
and $\bar\kappa$ is shown in Fig.~\ref{sl4}. The figure reveals several
universal features which appear to be independent
of the type of the {\sc sf} parametrization including:
({\it i\/}) strong dependence of $\bar{\varepsilon} (A)$ on
kinematics (shell effects); ({\it ii\/}) almost no
dependence of $\bar{\kappa} (A)$ on kinematics; ({\it iii\/})
clear surface ($\sim \frac{1}{A^{4/3}}$) dependence reducing
the dominant volume term ($\sim \frac{1}{A}$) in both
${\bar \varepsilon} (A)$ and  ${\bar \kappa} (A)$.

Indeed, the values of  $\bar{\varepsilon} (A)$  show characteristic
kinks close to double-(semi)magic $A$-numbers. These kinks
are magnified when all the calculated nuclei are used (no smoothing)
to compute $\bar{\varepsilon} (A)$,
but without affecting qualitatively the overall profile of
the curve. On the other hand, $\bar{\kappa}(A)$ is almost perfectly smooth
with barely visible traces of shell structure.
It confirms our earlier conclusion~\cite{[Sat03]}
that the gross features of the Skyrme
isovector mean potential can be almost
perfectly quantified by a smooth curve parametrized by
a small number of global parameters.

\begin{figure}[th]
\par
\centerline{\epsfig{file=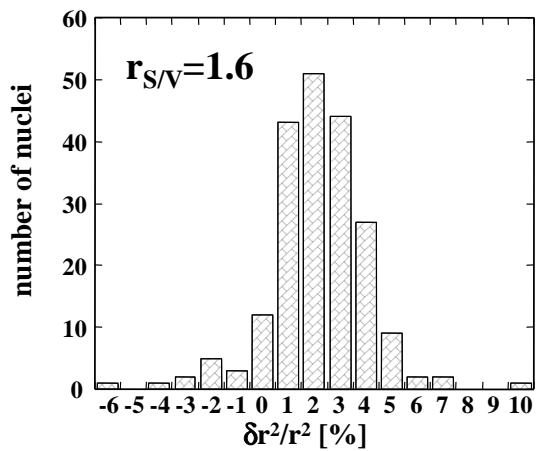,height=6cm,clip=}}

\caption[]{Histogram showing the deviation between neutron skin thickness
estimated using the simple hydrodynamical formula~(\ref{deltar}) taken
for $r_{S/V}=1.6$ and the results of microscopic~\shf~calculations.
Note, that the deviation is of the order of $\pm$3\%.
Note also, that the centroid of the histogram is slightly shifted
to the right what indicates that $r_{S/V}$ should be
slightly larger $r_{S/V}\sim 1.65$.
 } \label{skin}
\end{figure}


The~\shf~models yield~\cite{[Sat05c]} $r_{S/V}\sim 1.6$ in accordance with
the~\ld~ratio~\cite{[Mol95]}. There are several observables which
are strongly sensitive to $r_{S/V}$.  These include in particular
the static dipole polarizability ({\sc sdp})
and neutron skin thickness.
Using the hydrodynamical model of~\cite{[Lip82]} one can derive
simple expressions for both the {\sc sdp} and the neutron skin thickness.
It appears that
for $r_{S/V}\sim 1.6$  these simple expressions yield results
that are very consistent both with the data (for {\sc sdp}) and
the microscopic~\shf~calculations (neutron skin thickness),
see~\cite{[Sat05c]}. For example, the hydrodynamical formula for
the neutron skin thickness [in the lowest order expansion in $\frac{1}{A}$]
reads:
\begin{equation}\label{deltar} \frac{\delta r^2}{\langle r^2
\rangle}  \approx \frac{N-Z}{A} \left\{  1 + \frac{2}{3}
\frac{r_{S/V}}{A^{1/3}} - \ldots \right\}.
\end{equation}
The interesting feature of Eq.~(\ref{deltar}) is that it does not
depend on the bulk~\nse~coefficient but only on
the ratio $r_{S/V}$. This formula can  easily be cross-checked with our
microscopic \shf~calculations and the results are depicted in the form of
a histogram in Fig.~\ref{skin}.

\begin{figure}[th]
\par
\centerline{\epsfig{file=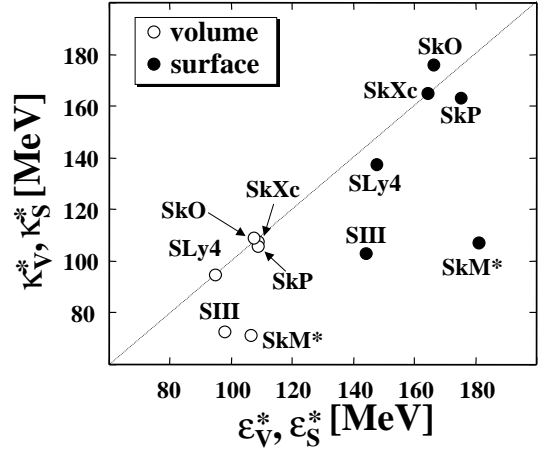,height=6cm,clip=}}

\caption[]{The correlation between the effective
mass scaled volume
$\varepsilon_V^\star$ and $\kappa_V^\star$ (open dots) and the surface
$\varepsilon_S^\star$ and $\kappa_S^\star$ expansion coefficients.
Note, that except for the SIII and SkM$^\star$ interactions, the expansion
coefficients are equal. } \label{e-k-star}
\end{figure}

The most striking result of our
analysis is the {\bf near-equality}
of $\bar{\varepsilon}^\star \approx \bar{\kappa}^\star$
[where $\bar{\kappa}^\star\equiv
\bar{\kappa} \frac{m_1^\star}{m} $ and $m_1^\star$ denotes
the standard isovectorial effective mass]
occurring for all modern parameterizations,
see Fig.~\ref{e-k-star}.  Indeed,
$\bar{\varepsilon}^\star$ differs from ${\bar\kappa}^\star$
only for old parameterizations like the SIII and SkM$^*$.
This result confirms the rather loose claims often
appearing in the textbooks that "{\it the kinetic energy\/}
[$\varepsilon_{FG}$] {\it and the isovector mean-potential contribute to the
$a_{sym}$ in a similar way\/}" is indeed correct but only after disregarding
non-local effects.
To our knowledge, it has never been discussed why this apparently independent
quantities should be similar.

\section{Summary}

\begin{figure}[th]
\par
\centerline{\epsfig{file=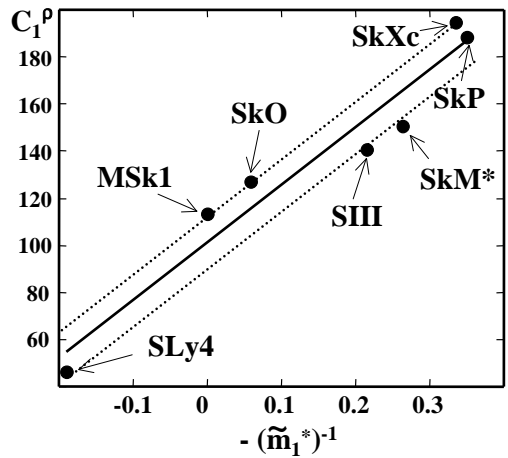,height=6cm,clip=}}

\caption[]{Correlation between $C_1^\rho$ and {\it kinematic\/} isovectorial
effective mass
$(\tilde m_1^\star)^{-1} \equiv \frac{2m}{\hbar^2}C_1^\tau\rho_0$
for different Skyrme force parameterizations. The $C_1^\rho$ and
$C_1^\tau$ are coupling constants defining
$C_1^\rho \rho_1^2$ and $C_1^\tau \rho_1 \tau_1$
isovector terms in the Skyrme local energy density functional.
 } \label{c1rho}
\end{figure}

The global mass dependence of the~\nse~strength $a_{sym}(A)$,
$\bar\varepsilon (A)$, and $\bar\kappa (A)$
is studied in detail within the~\shf~theory.
Our study enables us to establish the surface-to-volume ratio
of $a_{sym} (A)$, $r_{S/V}\approx 1.6$. This value is
in agreement with the~\ld~of Ref.~\cite{[Mol95]}
and is consistent with simple hydrodynamical estimates
for the {\sc sdp} and neutron skin thickness. Our study also reveals
an almost linear correlation between
$C_1^\rho$ and $({\tilde{m}_1^\star})^{-1}$, see Fig.~\ref{c1rho},
 and a striking
similarity between $\bar{\varepsilon}^\star \approx
\bar{\kappa}^\star$~\cite{[Sat05c]}.
The latter near equality indicates that {\it
the contribution to $a_{sym}$ due to the mean-level spacing and due to
the mean-isovector potential are similar\/} but only after
disregarding non-local effects.
Whether this is a fundamental property of the nuclear mean field
is an open question that requires further studies.

\section{Acknowledgments}
This work was been supported by Foundation for Polish Science (FNP),
the G\"oran Gustafsson Foundation, the Swedish Science Council (VR),
the Swedish Institute (SI), and the Polish Committee for Scientific
Research (KBN) under Contract No. 1~P03B~059~27.


\end{document}